\documentclass{ws-ijmpcs}

\newcommand{\beq}{\begin{equation}}
\newcommand{\eeq}{\end{equation}}
\newcommand{\bea}{\begin{eqnarray}}
\newcommand{\eea}{\end{eqnarray}}
\def\hsp{,\hspace{.7cm}}

\begin{document}

\markboth{Jarah Evslin}
{Challenges Confronting Superluminal Neutrino Models}

\catchline{}{}{}{}{}
%
%%%%%%%%%%%%%%%%%%%%%%%%%%%%%%%%%%%%%%%%%%%%%%%%%%%%%%%%%%%%%%%%%%%%

\title{Challenges Confronting Superluminal Neutrino Models
}

\author{JARAH EVSLIN}

\address{Theoretical Physics Center for Science Facilities\\
Institute of High Energy Physics, Chinese Academy of Sciences\\
YuQuanLu 19B, Beijing 100039, China\\
jarah@ihep.ac.cn}

\maketitle

\begin{history}
%\received{Day Month Year}
%\revised{Day Month Year}
\end{history}

\begin{abstract}
This talk opens the CosPA2011 session on OPERA's superluminal neutrino claim.  I summarize relevant observations and constraints from OPERA, MINOS, ICARUS, KamLAND, IceCube and LEP as well as observations of SN1987A. I selectively review some models of neutrino superluminality which have been proposed since OPERA's announcement, focusing on a neutrino dark energy model.  Powerful theoretical constraints on these models arise from Cohen-Glashow bremsstrahlung and from phase space requirements for the initial neutrino production.  I discuss these constraints and how they might be evaded in models in which the maximum velocities of both neutrinos and charged leptons are equal but only superluminal inside of a dense medium.

%The abstract should summarize the context, content
%and conclusions of the paper in less than 200 words. It should
%not contain any references or displayed equations. Typeset the
%abstract in 8 pt roman with baselineskip of 10 pt, making 
%an indentation of 1.5 pica on the left and right margins.

\end{abstract}

\ccode{PACS numbers: 11.30.Cp, 13.15.+g}

\section{Observations of Neutrino Velocities}

The OPERA\cite{opera} and MINOS\cite{minos} collaborations claim to have observed {\it{superluminal}} $\mu$-neutrinos with a fractional superluminality
\beq
\epsilon=\frac{v-c}{c} \sim 3\times10^{-5}. \label{opsup}
\eeq
In this talk I will not speculate on whether this claim is correct or incorrect.  In my opinion this determination can only be made by future experiments, and it will be made by future experiments perhaps within a few months, probably within a few years.  Instead I will ask the following question:  If neutrino superluminality is confirmed by future experiments, what does it teach us about physics?  In other words, what models are consistent both with OPERA's results, and with other known experimental and theoretical constraints?

%\scriptsize
%\begin{block}{OPERA}

\begin{table}%[ph]
\tbl{Superluminal Neutrino Velocity Observations and Bounds}
{\begin{tabular}{|c|c|c|}
\hline
\multicolumn{1}{|l|}{OPERA}& \multicolumn{2}{c|}{2009-2011}\\
\hline
Energy & Neutrinos & $ (v-c)/c$ \\
{ 10-50 GeV}& { 16,111 $\nu$'s (97\% $\nu_\mu$'2)} &
$2.48\pm0.28 \textrm{ (stat.)} \pm 0.30 \textrm{ (syst.)} \times 10^{-5}$\\
\hline
\multicolumn {1}{|l}{Distance:} & \multicolumn{2}{c|}{730 km from  CNGS (CERN) to OPERA (Gran Sasso)}\\ \hline
\hline
\multicolumn{1}{|l|}{MINOS}& \multicolumn{2}{c|}{May 2005-February 2006}\\
\hline
 Energy: 3 GeV & Neutrinos & $ (v-c)/c$ \\
{ (tail to 120 GeV)}& { $473\ \nu$'s ($93\%\ \nu_\mu$'s)} &
$5.1\pm 1.3$ (stat.)$\pm 2.6$ (sys.)$\times 10^{-5}$\\
\hline
\multicolumn {1}{|l}{Distance:} & \multicolumn{2}{c|}{734 km: Near Detector (FermiLab) to Soudan iron mine}\\ \hline
\hline
\multicolumn{1}{|l|}{Kamiokande II}& \multicolumn{2}{c|}{7:35 UT, February 23rd, 1987}\\
\hline
 Energy &
 Neutrinos &
$\nu$'s $\subset$ 13 sec.,  $\lesssim 3$ hrs before $\gamma$'s,  \\
 { 7.5-36 MeV}& { $12\ \bar\nu_e$'s} &
$ (v-c)/c<3\times 10^{-9}$ or $2\times 10^{-12}$\\
\hline
\multicolumn {1}{|l}{Distance:} & \multicolumn{2}{c|}{ 160,000 lys: Tarantula Nebula to Kamioka Observatory}\\ \hline
\hline
\multicolumn{1}{|l|}{Irvine-Michigan-Brookhaven}& \multicolumn{2}{c|}{7:35 UT, February 23rd, 1987}\\
\hline
 Energy & Neutrinos &   $\nu$'s $\subset$ 6 sec.,  $\lesssim 3$ hrs before $\gamma$'s,  \\
 20-40 MeV &  $8\ \bar\nu_e$'s & {  (v-c)/c$<3\times 10^{-9}$ or $2\times 10^{-12}$}\\
\hline
\multicolumn{1}{|c}{Distance} & \multicolumn{2}{c|}{ 160,000 lys: Tarantula Nebula to Morton-Thiokol salt mine}\\ \hline
\end{tabular}} \label{obstab}
\end{table}

The most obvious experimental constraint arises from neutrino observations of SN1987A.  We will begin by reviewing the MINOS, OPERA and SN1987A neutrino observations summarized in Table~\ref{obstab}.
OPERA and MINOS measured neutrinos created by colliding respectively 400 and 120 GeV protons into a graphite target.  This collision produces many positively charged pions and kaons which are focused and channeled into a decay pipe where they decay into an antimuon and a muon neutrino, which then flies 730 km through the Earth's crust to the detector.  It will be important in what follows that the density of the gas inside of both decay pipes is between $5\times 10^{-8}$ and $5\times 10^{-7}$ times that of the Earth's crust.  The resulting neutrinos have an average energy of 17 and 3 GeV respectively, but with long high energy tails.

Electron antineutrinos from SN1987A, which is about 160,000 light years away, were detected simultaneously at Kamiokande II\cite{k21987} (12), IMB\cite{imb1987} (8) and also Baksan\cite{bak1987} (5).  They were detected at 7:35 in the morning universal time, 3 hours before visual light from the supernova was photographed at 10:40\cite{luce}.  While this photograph establishes an upper bound on the time at which the supernova became visually brighter, the lower bound is established by photographs exposed between 1:20 and 2:25\cite{garrison} in which the supernova does not appear.   The most often cited lower bound is 9:35, when an experienced amateur astronomer did not notice anything unusual while observing stars a quarter of a degree away.  It has been claimed\cite{wampler} that he would have noticed had the star brightened by more than 3 magnitudes.  The 10:40 observation sets a maximum fractional superluminality
\beq
\epsilon<3\times 10^{-9}.
\eeq
However, considering both the potential delay of the light and the possibility that light from the supernova may have arrived before 7:35, subluminal neutrino velocities, beyond those predicted by relativity, are far from excluded.

A tighter superluminality bound can arise from the fact that despite the wide range in neutrino energies, varying from 7.5 to 40 MeV, all of the neutrinos were observed within a 13 second window, with no obvious correlation between the energies and times of arrival.  Therefore if the superluminality depends strongly on the energy, for example by a power law with a nonzero integral power, then the bound is significantly tightened, to $\epsilon<2\times 10^{-12}$.  This can be further tightened by observing that the neutrino arrival rate was exponentially decreasing during these 13 seconds.  In fact at each of these 3 experiments, more than half of the neutrinos arrived during the first two seconds of the event (9, 5 and 3 respectively.)  Therefore if the superluminality has a power law dependence on the energy one can obtain the very strong bound
\beq
\epsilon<4\times 10^{-13}. \label{stretta}
\eeq

\section{Exploiting differences between OPERA and SN1987A neutrinos}

Clearly the fractional neutrino superluminality (\ref{opsup}) reported by MINOS and OPERA is inconsistent with both bounds imposed by neutrinos from SN1987A.   This implies that any successful model of neutrino superluminality must exploit a feature of these experiments which distinguishes OPERA and MINOS neutrinos from SN1987A neutrinos.  There are four obvious\footnote{Other distinguishing features include background temperatures\cite{matone} as well as the directions of travel which always differed by more than 90 degrees in the frames of the Earth, galaxy and CMB.} distinguishing features:
\begin{itemlist}
 \item {\bf{Lepton number}}

The observed SN1987A neutrinos were all antineutrinos.  By comparison, OPERA and MINOS neutrinos are nearly all honest neutrinos.  To exploit this distinction a model would involve a very bizarre CP violation.  However, bizarre CP violation may already be required to explain the MiniBooNE anomaly\cite{miniboone}.  $2\%$ of neutrinos observed by OPERA are indeed antineutrinos, although it may not always be possible to determine which $2\%$ these are.  This means more than 300 neutrinos.  It may therefore be possible to test this hypothesis with current data, at least at the 1 or 2$\sigma$ level it should be possible to determine whether antineutrinos are also superluminal.

 \item {\bf{Flavor}}

One may attempt to reconcile these observations by allowing only muon neutrinos to be superluminal.  Neutrino oscillation experiments measure the difference in the dispersion relations of different flavors of neutrinos.  KamLAND\cite{kamland} has observed reactor electron antineutrinos between 1.8 and 9 MeV.  By transforming their results, one can determine that these neutrinos oscillate with a characteristic oscillation length of order 10 km.  On the other hand, the above hypothesis, with a simple dispersion relation, yields an oscillation length of order nanometers, and even less for more complicated dispersion relations\cite{us}.  Therefore in this case KamLAND would not directly observe the oscillations, but rather would observe an energy-independent antineutrino survival probability.  This is disfavored but certainly not excluded by their results.

 \item {\bf{Energy}}

OPERA neutrinos are about 1000 times more energetic than SN1987A neutrinos.  Therefore, perhaps only high energy neutrinos are superluminal? The simplest possibility would be if the superluminality depended on the neutrino energy $E$ via a power law.  Then one could fit OPERA data and satisfy supernova constraint (\ref{stretta}), while disregarding the low confidence MINOS data, with
\beq
\epsilon\propto E^\alpha \hsp \alpha>2.6 .
\eeq

In anticipation of such models, OPERA attempted to determine the energy dependence of the neutrino superluminosity.  They cannot determine the energies of many of their neutrinos, as they decay outside of the detector.  However 5,489 of their neutrinos interact inside of the detector via charged currents.  OPERA divided these into two groups, those with energy less than and greater than 20 GeV, and measured the superluminality of both.  The first group had an average energy of 13.9 GeV and $\epsilon=(2.2\pm 0.8\pm 0.3)\times 10^{-5}$ while the second had an average energy of 42.9 GeV and a superluminality of $\epsilon=(2.7\pm 0.8\pm 0.3)\times 10^{-5}$.   A power law fit therefore implies not only a 1.7$\sigma$ discrepancy with MINOS data, but also a similar discrepancy with the 13.9 GeV group.  However, unlike the main OPERA result most of the error in this analysis is statistical, and so this poor fit may improve with future runs.  Of course, one may obtain a perfect fit with a more complicated dispersion relation\cite{strumia,miaoli}.   As we will explain, all such models are probably excluded for other reasons.

 \item {\bf{Location}}

MINOS and OPERA neutrinos traveled almost entirely through solid rock, while SN1987A neutrinos traveled almost entirely through the interstellar medium.  Therefore, perhaps neutrinos are only superluminal while traveling through dense media?
Such an effect may be realized in a model if, for example, the background stress tensor or baryon density couples to isolated objects\cite{tianjun} or to a tensor\cite{alex}, vector\cite{ellis} or scalar\cite{kehasias,us} field which in turn is coupled kinetically to the neutrino field, modifying the effective metric seen by neutrinos and in particular spontaneously breaking Lorentz symmetry.  While the models just cited were all found independently during the week after OPERA's announcement, for illustration we will now describe a model based on neutrino dark energy\cite{xinmindark}.

\end{itemlist}

\section{Neutrino dark energy model}
Applying the usual logic of effective field theories, one may write down all of the interactions of the dark energy scalar field $\Pi$ with the neutrino.  The lowest dimensional terms in the effective action which will be relevant are
\beq \Delta\mathcal{L}=\frac{1}{2}\left(
ia_\mu\bar{\nu}\partial^\mu\nu+ic_{\mu\nu}\bar{\nu}\gamma^\mu\partial^\nu\nu
-d_{\mu\nu\rho}\bar{\nu}\gamma^\mu\partial^\nu\partial^\rho\nu\right) \label{nuaz}
\eeq 
where $a$, $c$ and $d$ are tensors constructed from derivatives of $\Pi$, for example via
\beq c_{\mu\nu}=-\frac{b\langle
v\rangle^2}{2}\langle\partial_\mu\partial_\nu\Pi  \rangle
\eeq
where $v$ is the Higgs VEV, which appears so as to simplify the manifestly SU(2)-invariant form of the coupling.  The constant $b$ is a free parameter.  We will ignore the time dependence of $\Pi$, and we not impose that it actually be responsible for the dark energy of our universe, but hope that it is.  The corresponding fractional superluminality on the $x-y$ plane is
\beq
\frac{v-c}{c}\simeq \frac{a_x^2}{2}\cos^2\theta+\left(c_{xx}\cos^2\theta+c_{yy}\sin^2\theta\right)
+2E\cos\theta\left(d_{xxx}\cos^2\theta+3d_{xyy}\sin^2\theta\right) 
\eeq
where $\theta$ is the angle between the neutrino's momentum and the $x$ axis.  

In this model one can consider a very simple coupling of $\Pi$ to the trace of the background stress tensor $T$
 \beq
\L_\Pi=-\frac{1}{2}\partial_\mu\Pi\partial^\mu\Pi+4\sqrt{3\pi G_N}\Pi T. \label{paz}
 \eeq 
Fifth force constraints can be satisfied by simultaneously scaling the coefficient of T and also $b$, as only the product appears in the superluminality.  The tensor $d$ couples to a higher dimensional operator, while $a$ only appears quadratically in the velocity, therefore by the usual logic of effective theories both may be ignored.  Then the theory is completely specified.  One can find the $\Pi$ field sourced by the Earth, the Sun and the Galaxy and so calculate the superluminosities of OPERA, MINOS and SN1987A neutrinos.  This has been done\cite{us}, fixing the single parameter $b\sim (6\ {\rm{keV}})^{-5}$ easily fits them all.

\section{Two difficult challenges}
In the time that has passed since these models were submitted to the ar${\rm\chi}$ive, two critical observations have been made.  These theoretical observations are sometimes believed to be arguments against the results of the experiment OPERA itself.  Of course a theoretical argument cannot disprove an experiment, only a failure to repeat the experiment by another group can do that.  However a theoretical argument can rule out a model, if the assumptions of the argument are satisfied by the model.  In the case at hand, the assumptions of the arguments are not satisfied in general but they are satisfied to a very good approximation by nearly all superluminal models that have appeared thus far.  Those that escape them very likely escape only because they have not yet provided enough details to be falsified.  In this section I will briefly review these two challenges, and speculate on how they may be avoided.

\subsection{Cohen-Glashow bremsstrahlung}

Cohen and Glashow have argued\cite{cohenglashow} that superluminal neutrinos lose energy due to the bremsstrahlung process
\beq
\nu\rightarrow \nu+e+\bar e.
\eeq
In particular, they claim that as a result neutrinos traveling 730 km will likely end their journey with less than 12.5 GeV.  This would not necessarily be a disaster for OPERA, after all when they divided their neutrinos into high energy and low energy bins, the high energy neutrinos were only superluminal by about 2 standard deviations.  However it is in conflict with the results of the ICARUS experiment, which in this case would have seen the nucleated electron positron pairs and did not\cite{icarus}.

Their argument rests on four assumptions:
\begin{romanlist}[(ii)]
\item Superluminality arises entirely from a $p^2$ term in the dispersion relation $E^2(p)$.
\item In their asymptotic final states, the electron and positron are noninteracting and on-shell in the vacuum.
\item The 4-momentum of the initial neutrino is the sum of those of the three final particles.
\item Electrons cannot be superluminal.
\end{romanlist}

Therefore any model of neutrino superluminality must violate at least one of these assumptions, and must violate it to a significant extent that not only can enough OPERA neutrinos arrive, but also ICARUS will not observe these pairs.  The first of these conditions may be satisfied with a severe modification of the dispersion relations\cite{disp}, but it seems unlikely that any such choice satisfies the second over the entire range of energies of OPERA neutrinos.  Certainly such a possibility has not yet been demonstrated.

The second and third assumptions are incorrect in any model.  Indeed, as MINOS and OPERA neutrinos travel nearly entirely through solid rock, there is no asymptotic state in which the electrons or positrons could move at relativistic speeds and some 4-momentum will certainly be transfered to the rock as an effect of this friction.  Furthermore, in theories in which the superluminality is caused by extra fields, these fields will also generically exchange 4-momentum with the electron-neutrino system.  No calculation of the magnitudes of these effects is available, but dimensional analysis suggests that they are not sufficient to render MINOS and OPERA results compatible.  Some models exhibit a more extreme violation of (iii) by modifying momentum conservation throughout the universe\cite{smolin} or even by modifying the relevant notion of momentum\cite{finsler}.

We will focus on assumption (iv).  Cohen and Glashow demonstrated that the threshold neutrino energy for their process is
\beq
E_\nu=\frac{2 m_e}{\sqrt{v_\nu ^2-v_e^2}}
\eeq
where $v_\nu$ and $v_e$ are the maximum neutrino and electron velocities.  SU(2) gauge-invariance already suggests that above the electroweak scale these velocities be equal, and so the threshold energy will be infinite and the process will not occur.  We will now consider the more extreme case in which the neutrino and electron superluminality are always equal.  With this assumption there will no longer be neutrino bremsstrahlung, but electrons may now be superluminal.

Is electron superluminality ruled out by experiment?  If the neutrinos are only superluminal inside of dense matter, then electrons would also only be superluminal inside of a dense medium.  As the media at our disposal all interact electromagnetically, there are no experiments in which one may hope to saturate this bound, and so no direct experimental constraints on electron velocity in a high density region.  Thus if a model only predicts neutrino superluminality inside of a dense medium, at least direct constraints on electron velocity are avoided\footnote{However changes in phase space for interactions involving leptons may have significant consequences in dense regions such as stellar cores, white dwarves and in the very early universe.}.

While this may or may not be the case for models in which superluminality is caused by interactions with discrete objects\cite{tianjun}, unfortunately in the case of the models above in which superluminality is caused by a very light field, the superluminality extends well beyond the Earth's crust, into regions in which relativistic electrons have been observed.   These models therefore need to be tested against experiments that measure the maximum electron velocity not only inside of solid rock, but also nearby.  

The two strongest constraints on electron velocities at these energy scales come from synchrotron radiation at LEP\cite{LEP} at OPERA energies and in the Crab Nebula\cite{crab} at hundreds of TeV.  Both of these imply the constraint $\epsilon\lesssim 10^{-15}$.   The electrons in both cases were in very diffuse environments, with densities of order $10^{-17}$ and $10^{-20}$ times that of rock respectively.   Therefore, these constraints on electron velocities, which become constraints on neutrino velocities if $v_\nu=v_e$, are compatible with OPERA and MINOS' $\epsilon=3\times 10^{-5}$ inside of rock if, for example, the superluminality scales with the density $\rho$ as a power law
\beq
\epsilon\propto \rho^\alpha\hsp \alpha>0.6. \label{colim}
\eeq
Unfortunately, since in the above spontaneous symmetry breaking models the superluminality is caused by effectively massless carriers, there is no local relation between $\epsilon$ and $\rho$ and so these models are likely to be ruled out by Cohen and Glashow's argument.  

One instead requires a massive field, with a mass of at least the inverse radius of the LEP tunnel, or at least a field which acquires a chameleon mass\cite{cham}.  This would then give a field whose value depends only upon the local density as desired.  However its gradients would then be too sharp to use in a model like the neutrino dark energy model above.  One would need to consider either a massive spin 2 model, at least $10^6$ times more massive than that of Dvali and Vikman, or else a neutrino dark energy model in which the neutrino is kinetically coupled to the product of a roughly spatially homogeneous quintessence scalar $Q$ whose value provides a universal time and a density dependent massive scalar $\Pi$
\beq c_{\mu\nu}=-\frac{b\langle
v\rangle^2}{2}\langle f(\Pi) \partial_\mu\partial_\nu Q  \rangle . \label{nuovo}
\eeq
Here the power law superluminality (\ref{colim}) follows if $f(\Pi)=\Pi^\alpha$.

\subsection{Meson decay kinematics}

Another strong constraint on superluminal neutrino models arises from the creation of the neutrino itself, in meson decay at CNGS and NuMI.   If the meson is not also superluminal (which indeed it is in some models) then there is only available phase space for the pion decays if\cite{kin}
\beq
E_\nu<(m_\pi-m_\mu)\sqrt{1+\frac{1}{2\epsilon}}.
\eeq
For OPERA this is only about 5 GeV.  Of course even if one evades the bound, this effect nonetheless causes distortion in the neutrino spectra if one is too close.  Recall that these mesons decay in a vacuum tube with a density of less than $10^{-6}$ times that of rock.  Therefore the density dependence (\ref{colim}) yields $\epsilon<10^{-8}$ and so automatically also satisfies this constraint.

Stronger constraints arise from IceCube neutrinos\cite{icecube}, which travel across the Earth with an anomalous loss of at most $20\%$ and have energies up to and above 400 TeV.  Of course, there is no evidence for neutrino superluminality at such high energies.  Cohen-Glashow bremsstrahlung is again no problem if the electron and neutrino maximum velocities are equal, but one does need to worry about the kinematics of the decay which led to the original neutrino production.  Near 100 TeV most of these atmospheric neutrinos result from prompt decays of charmed mesons\cite{prompt}, which are much more massive and so yield weaker kinematic energy bounds.   Therefore one need only extend the maximum neutrino energy to about 100 TeV.  

In popular models these neutrinos are created in the upper atmosphere, where the density is about $10^{-5}$ times that of rock.  Therefore if indeed the neutrino superluminality in rock is still of order $10^{-5}$ at these energies, then either these neutrinos are created much higher in the atmosphere than is thought, or else $\alpha=1$ is excluded.  More precisely, at $\alpha=1$ pion decays cannot generate 100 TeV neutrinos below about 55 km, and kaon decays below about 25 km, but of course an appreciable distortion in the spectrum persists to higher elevations due to the reduced phase space.  $\alpha=2$ comfortably avoids this bound, but may lead to $1\%$ level superluminality in the solar core, which would affect the kinematics of solar fusion.  Are such models then already ruled out?

\subsection{Evading the constraints}
In summary, if the maximum neutrino and charged lepton velocities are everywhere equal then there will be no Cohen-Glashow bremsstrahlung.  Constraints on superluminal neutrinos from meson decay kinematics and SN1987A as well as constraints on electron superluminality from synchrotron and Cherenkov radiation all arise from extremely low density regions.  Therefore all of these constraints are avoided if the local value of the superluminality is tied to the local density, as in the neutrino dark energy model of Eqs.~(\ref{nuaz}), (\ref{paz}) and (\ref{nuovo}) with a sufficiently massive scalar $\Pi$.  If on the other hand the superluminality is controlled by a very light field, as in previous models, then it will not be a local function of the density and so these models are likely to be ruled out.  If superluminality extends up to the higher IceCube energies, then in addition to Cohen and Glashow's bremsstrahlung process, neutrinos may also lose energy due to $\nu\rightarrow\nu+\pi^0$ and at one-loop $\nu\rightarrow\nu+\gamma$.  The corresponding bounds may may well imply that all particles need to have the same superluminal maximum velocity, a feature of these models which could potentially be in conflict with big bang nucleosynthesis.

\section{When will we know?}
A number of theoretical arguments have suggested that the neutrino superluminality observed at OPERA and MINOS is inconsistent.  In this talk, we have described how models can be constructed that may potentially evade these arguments.  If it turns out the OPERA and MINOS have indeed observed superluminal neutrinos, such loopholes will need to be built into any physical theory of neutrino physics.  This leads us to ask, when will we know whether OPERA and MINOS are right?

OPERA's conclusion is particularly suspicious because the superluminality is energy independent.  This has led a number of authors to suspect a systematic error, which in most scenarios simply shifts the neutrino travel time.  The largest source of systematic errors cited by the OPERA collaboration is indeed of this kind.  The measured neutrino velocity is simply the distance that they have traveled, divided by the time that has passed since they were formed.  Of course the time at which they were formed is not known and so one uses the time at which the protons passed a detector and attempts to correct for the relevant timelags.  But the problem is that the protons are sent in continuous extractions which last about 11,000 nanoseconds, and so it is impossible to determine which proton corresponds to which neutrino, one instead performs an extensive statistical analysis.  The measurement of the original proton timing is responsible for most of the systematic error estimated by the OPERA collaboration\cite{opera}, and furthermore many have claimed that this source of error has been grossly underestimated by the collaboration.

The good news is that last week the CNGS began sending OPERA 1-2 nanosecond extractions, separated by 500 nanoseconds.  This will allow an unambiguous identification of which proton corresponds to which neutrino.  Unfortunately, this will stop on November 21st when the SPS, which is accelerating the protons, needs to accelerate lead ions for ALICE for a month.  Nonetheless, this new run provides an opportunity to falsify OPERA's claims in the next few months.  

Of course there are many possible sources of systematic error, and so one can never place one's confidence in a single experiment, especially one with a constant baseline.  Before the OPERA result should be believed, it must be repeated. 
MINOS' evidence for superluminality has been unconvincing for 4 years because of their large systematic errors.   These systematic errors are in part different from those of OPERA.  For example, it has two neutrino detectors and so does not need to rely on the proton timing, it can directly measure the time that the neutrino beam arrives at each detector.  The  MINOS systematic error is dominated by a large uncertainty in the delay time of the optical cable that transports the GPS timing down to the detectors.  An accurate measurement of this delay time can be used to reevaluate the past 6 years of data, substantially reducing the systematic uncertainties.  Such a rerelease of old data is indeed planned in the next 4-6 months, with new data perhaps already next year.

Thus it seems as though in perhaps a few months, probably within a few years, we will know whether or not these superluminality claims are correct.  However for a convincing demonstration that one is indeed observing an effect caused by the neutrinos speed, one must show that the anomalous travel time is indeed proportional to the baseline.  As OPERA and MINOS have essentially identical baselines, they cannot test this thesis.  Perhaps T2K can, but it's baseline is so short that its timing will first need to be significantly improved.  A definitive positive answer is then only attainable from proposed longer baseline experiments, for example from Japan to South Korea or from CERN to Finland.  On the other hand a definitive negative answer may arrive at any time.

\section* {Acknowledgement}

\noindent I have benefited beyond any reasonable measure from discussions with Emilio Ciuffoli, Andy Cohen, TianJun Li, Alex Vikman and XinMin Zhang and from support by the Chinese Academy of Sciences
Fellowship for Young International Scientists grant number
2010Y2JA01. 

\end{document}